\documentclass[reprint,amsmath,amssymb,aps,floatfix,showkeys,showpacs,
]{revtex4-1}

\usepackage[T1]{fontenc}
\usepackage[utf8]{inputenc}

\usepackage{graphicx,colordvi,longtable,xcolor,color}
\usepackage{dcolumn}
\usepackage{bm}

\newcommand{\cn}{\,{\sf cn}}
\newcommand{\sn}{\,{\sf sn}}
\newcommand{\dn}{\,{\sf dn}}
\newcommand{\sech}{\,{\sf sech}}
\newcommand{\tnh}{\,{\sf tanh}}

\begin{document}

\title{ Exact solitonic and periodic solutions of the extended KdV equation
}
\thanks{Dedicated to the memory of Dr John Dougherty  (1935 - 2015).}

\author{Eryk Infeld} \email{Eryk.Infeld@ncbj.gov.pl}
\affiliation{National Centre for Nuclear Research, Hoża 69, 00-681 Warszawa, Poland}

\author{Anna Karczewska}
 \email{A.Karczewska@wmie.uz.zgora.pl}
\affiliation{Faculty of Mathematics, Computer Science and Econometrics\\ University of Zielona G\'ora, Szafrana 4a, 65-246 Zielona G\'ora, Poland}

\author{George Rowlands} \email{G.Rowlands@warwick.ac.uk}
\affiliation{Department of Physics, University of Warwick, Coventry, CV4 7AL, UK}

\author{Piotr Rozmej}
 \email{P.Rozmej@if.uz.zgora.pl}
\affiliation{Institute of Physics, Faculty of Physics and Astronomy \\
University of Zielona G\'ora, Szafrana 4a, 65-246 Zielona G\'ora, Poland}

\date{\today}

\begin{abstract}
The KdV equation can be derived in the shallow water limit of the Euler equations. Over the last few decades, this equation has been extended to include both higher order effects (KdV2) and an uneven river bottom. Although this equation is not integrable and has only one conservation law, exact periodic  and solitonic solutions exist for the even bottom case. The method used to find them assumes the same function forms as for KdV solutions. KdV2 equation imposes more constraints on parameters of solutions. For soliton case KdV2 solution occurs for particular ratio of wave amplitude to water depth only. For periodic case
physically relevant solutions are admissible only for two narrow intervals of elliptic parameter $m$.
For a range of $m$ near one the cnoidal waves are upright as expected, but are inverted in  $m$ region close to zero.
Properties of exact solutions of KdV and KdV2 are compared.

\end{abstract}



\maketitle

\section{Introduction}
One hundred and seventy years ago, Stokes pointed out that waves
described by nonlinear models can be periodic \cite{Stokes}. Although several related results followed, it took half a century before the Korteveg - de Vries equation became widely known \cite{KdV}. A competitive equation, Boussinesq, was formulated
in 1871. It is also the theme of several recent papers \cite{BBM,Bona81}.
Another direction research
has gone in is including perpendicular dynamics in KdV, e.g.~\cite{IRS99}.

The KdV equation is one of the most ubiquitous physical equations. It consists of the mathematically simplest possible terms representing the interplay of nonlinearity and dispersion. This simplicity may be one of the reasons for success. Here we investigate this equation improved as derived  from the Euler inviscid and irrotational water equations. 

Just as for conventional KdV, two small parameters are assumed: wave amplitude/depth $(a/H)$ and depth/wavelength squared $(H/l)^2$. These dimensionless expansion constants are called $\alpha$ and $\beta$. We take expansion one order higher. The new terms will then be of second order. This procedure limits considerations to waves for which the two parameters are comparable. Unfortunately some authors tend to be careless about this limitation.

The next approximation to Euler's equations for long waves over a shallow riverbed is ($\eta$ is the elevation above a flat surface divided by $H$)
\begin{align} \label{kdv2}
 \eta_t & +  \eta_x + \frac{3}{2} \alpha\,\eta\eta_x+ \frac{1}{6}\beta\, \eta_{3x}\\ & -\!\frac{3}{8}\alpha^2\eta^2\eta_x  \!+\!
  \alpha\beta\!\left(\frac{23}{24}\eta_x\eta_{2x}\!+\!\frac{5}{12}\eta\eta_{3x}\! \right)\!+\!\frac{19}{360}\beta^2\eta_{5x} =0 .  \nonumber
\end{align} 

In (\ref{kdv2}) and subsequently we use low indexes for derivatives $\left(\eta_{nx}\equiv\frac{\partial^n \eta}{\partial x^n} \right)$. 
This second order equation was called by Marchant and Smyth \cite{MS90,MS96} the \textit{extended KdV}. It was also derived in a different way in \cite{BS13} and \cite{KRR14,KRI14}. We call it {\bf KdV2}. It is not integrable. 
Not only  is KdV2 non integrable, it only seems to have one conservation law (volume or mass) \cite{KRI15,SV16}. A simple derivation of adiabatically conserved quantities can be found in \cite{KRIR17}.

Many authors, e.g. \cite{Fokas,Dull2001}, argue that equations  like (\ref{kdv2}) can be transformed to an asymptotically equivalent integrable form. The asymptotic equivalence means, that solutions of these equations converge to the same solution when small parameters tend to zero. This approach was first introduced with \emph{near-identity transformation} (NIT) by Kodama \cite{Kodama} and then used and generalized by many others, e.g. \cite{Hiraoka,GrimPel,Dull2004}. However, NIT is an approximation in which terms of higher order are neglected.  Therefore, 
for finite values of small parameters $(\alpha,\beta)$, solutions of NIT-transformed integrable equation are not the same as  {\bf exact} solutions.

Exact solitonic solutions of KdV2 were found by us in \cite{KRI14}. In the main part of the paper we derive exact periodic solutions to (\ref{kdv2}).

\section{KdV solutions}
For further discussions of new properties of KdV2 solutions we first briefly remind KdV solutions.

\subsection{Standard approach}

KdV equation is given by the first line of (\ref{kdv2}). For standard approach to derive exact solutions, see, e.g.\ \cite{Whit,Ding}. Here we will remind only some steps of these derivations and the final results.
Introducing new variable $\xi=x-ct$, where  $c = 1 + \alpha c_1$ and dividing KdV equation by $\alpha$ one obtains an ODE equation
\begin{equation} \label{ode}
- c_1 \eta_\xi +\frac{3}{2} \eta \eta_\xi + \frac{1}{6}\frac{\beta}{\alpha}\eta_{3 \xi} = 0.
\end{equation}
Integration gives ($r$ is an integration constant)
\begin{equation}\label{odeI1}
- c_1 \eta + \frac{3}{4} \eta^2 + \frac{1}{6} \frac{\beta}{\alpha} \eta_{2 \xi} = \frac{1}{4} r.
\end{equation}
Then multiplication by $\eta_{\xi}$ and  next integration yields
\begin{equation} \label{odeI2}
\frac{1}{3}\frac{\beta}{\alpha}  \left(\eta_\xi \right)^2  = - \eta^3 + 2 c_1 \eta^2 + r \eta + s =: f\left( \eta \right),
\end{equation}
where $s$ is another integration constant. Then if solutions are such that 
$\eta(\xi)\to 0$ when $\xi\to \pm\infty$ then $r=s=0$. In this case $f(\eta)=\eta^2(2c_1-\eta)$ and integration of (\ref{odeI2}) is easily  obtained with hyperbolic functions leading finally to single soliton solution
\begin{equation}\label{1sol}
\eta(x,t)=\sech^2\left[ \sqrt{\frac{3\,\alpha}{4\,\beta}} \left(x- t \left(1+\frac{\alpha }{2}\right)\right)\right].
\end{equation}
 Note that the amplitude of this soliton solution is equal~1.

The path to obtain exact periodic solutions is much more involved. The most detailed discussion of this problem is contained in \cite{Ding}. Below, we remind only few important steps and formulas. 
In general, integration constants can be nonzero. Then, assuming that $\eta_1<\eta_2<\eta_3$ are roots of polynomial $f(\eta)$, the polynomial can be written as 
\begin{equation}
f(\eta) = - (\eta - y_1)(\eta - y_2)(\eta - y_3).
\end{equation}
Real-valued solutions are possible when $-\eta_2\le \eta\le \eta_1$, only.
Then solution of (\ref{odeI2}) can be found in the form
\begin{equation}\label{rcs}
\eta(\xi) = \eta_1 \rm{cos}^2 \chi(\xi) - \eta_2 \rm{sin}^2 \chi(\xi).
\end{equation}
Then (\ref{odeI2}) takes form
\begin{equation} \label{odeI3}
\frac{4\beta}{3\alpha}  {\chi_\xi}^2 = (\eta_1 + \eta_3) - (\eta_1 + \eta_2) \rm{sin}^2 \chi.
\end{equation}
Denoting $m=\frac{\eta_1+\eta_2}{\eta_1+\eta_3}\in [0,1]$ and $\Delta^2=\frac{4\beta}{3\alpha(\eta_1+\eta_3)}$ one obtains from (\ref{odeI2})
\begin{equation}\label{delch}
\Delta^2 \chi_\xi^2 = 1 - m \rm{sin}^2 \chi.
\end{equation}
Integration yields
\begin{equation}\label{cdelch}
\frac{1}{\Delta}\int_0^\xi d\hat{\xi}= \mp \int_0^\chi \frac{d \hat{\chi}}{\sqrt{1 - m \rm{sin}^2 \hat{\chi}}} \quad \Longrightarrow \quad \pm \frac{\xi}{\Delta} = F(\chi | m),
\end{equation}
where $F(\chi | m)$ is the incomplete elliptic integral of the first kind.
Since the inverse functions are
\begin{equation} \label{inv}
\cos \chi = \cn \left(\frac{\xi}{\Delta} | m \right), \qquad
\sin \chi = \sn \left(\frac{\xi}{\Delta} | m \right)
\end{equation}
then from (\ref{odeI3})  solution is obtained in the form
\begin{equation}\label{cn2Ding}
\eta(\xi)= - \eta_2 + (\eta_1 + \eta_2) \cn^2 \left(\frac{\xi}{\Delta} | m \right).
\end{equation}
In next steps Dingemans \cite{Ding} stresses three conditions which allows him to express  $\eta_1,\eta_2,\eta_3$ through physical quantities. Two of these conditions come from definitions of dimensionless variables. Since distance $x$ has been made by dimensionalization with the wavelength, then dimensionless wavelength should be equal to 1. Dimensionalization of vertical variable has been made with $H$ so dimensionless amplitude should be  equal to 1, as well (as it came automatically in the case of soliton solution (\ref{1sol})).
The third condition requires that the mean free surface elevation should coincide with still water surface.

\subsection{Algebraic approach}

From theory of nonlinear differential equation it is known, see, e.g. \cite{KhSa15}, that for some classes of such equations exact solutions should exist in forms of either hyperbolic or Jacobi elliptic functions. It appears that both KdV and KdV2 equations belong to these classes. Therefore one can directly look for solutions of these equations assuming a particular form of solution. Our main goal is to find exact solutions of KdV2 equation.
In order to introduce the reader to the algebraic approach we begin with much simpler KdV case.

\subsubsection{Single soliton solution}

Soliton solution is assumed as (amplitude is set equal~1)
\begin{equation}\label{asumkdv1}
\eta(x,t) = \sech^2[B(x-vt)]= \sech^2(By),
\end{equation}
where $y=x-vt$. Substitution (\ref{asumkdv1}) into KdV gives
\begin{equation}\label{akdv}
\frac{1}{3} B \tnh (B y) \sech^2(B y) 
 \left[\ G_0 + G_1 \sech^2(B y)\right]=0. 
\end{equation}
Equation (\ref{akdv}) is valid for any argument only when simulaneously
\begin{align}\label{G0}
G_0 & = 6 v-6 -4 \beta B^2 =0, \\ \label{G1}
G_1 & =12 \beta  B^2-9 \alpha =0. 
\end{align}
This gives immediately
\begin{equation}\label{Bv}
B^2=\frac{3\,\alpha}{4\,\beta}, \qquad v=1+\frac{\alpha}{2}
\end{equation}
and solution coincides with (\ref{1sol}).

{\sf Remark: \it
It is clear from (\ref{Bv}) that solutions exist for arbitrary parameters $\alpha,\beta$, provided both are small. Waves are unidirectional since $v>1$.}

\subsubsection{Periodic solution}

In this case solution is postulated in the form of cnoidal wave
\begin{equation}\label{cnoid}
\eta(x,t) = \cn^2[B(x-vt),m]+D.
\end{equation}
[Equivalently, instead of Jacobi elliptic $\cn$ function, $\dn$ or $\sn$ Jacobi elliptic functions can be used.]

Then, substitution of (\ref{cnoid}) into KdV yields equation analogous to (\ref{akdv})
\begin{equation}\label{pkdv}
(-2B\,\cn\sn\dn)\left[G_0 +G_1 \cn^2 \right] =0.
\end{equation}
So, there must be
\begin{align} \label{g0}
G_0 &= 1-v+\frac{3}{2}D\alpha- \frac{2}{3}B^2\beta+\frac{4}{3}m\,B^2\beta=0,  \\ \label{g1}
G_1 &=\frac{3}{2}\alpha - 2m \,B^2\beta=0. 
\end{align}
Eq. (\ref{g1}) implies 
\begin{equation}\label{B2pkdv}
 B^2=\frac{3\,\alpha}{4\,\beta}\frac{1}{m}.
\end{equation}
Volume conservation condition (details will be explained later) determines
\begin{equation}\label{vccon}
D = - \frac{1}{m} \left[\frac{E(m)}{K(m)}+m-1\right].
\end{equation}
In (\ref{vccon}), $E(m)$ and $K(m)$ are the complete elliptic integral and  the complete elliptic integral of the first kind, respectively.  Then from (\ref{g0}) one has
\begin{equation}\label{vpkdv}
v=1 - \frac{\alpha}{2}\left[3\frac{E(m)}{K(m)}+m-2\right]\frac{1}{m}\equiv 
1- \frac{\alpha}{2}\,\frac{\text{EK}(m)}{m}.
\end{equation}
where an auxiliary function $\text{EK}(m)=\left[3\frac{E(m)}{K(m)}+m-2\right]$ was introduced. Function $\text{EK}(m) $ decreases monotonically from 1 for $m=0$ to -1 for $m=1$ and has the root at $m_s\approx 0.9611495$.

\begin{figure}[tbh] 
\begin{center}
\resizebox{0.9\columnwidth}{!}{\includegraphics{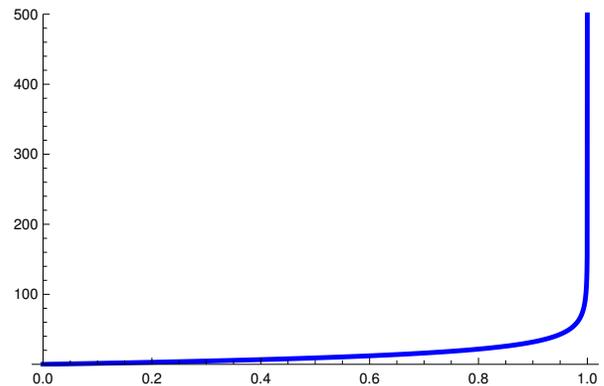}} 
\end{center}
\vspace{-5mm}
\caption{Ratio $\frac{\alpha}{\beta}$ (\ref{lpkdv}) as function of $m$.} \label{aDb}
\end{figure}

However, there is relation between parameters $\alpha,\beta$ and $m$ which limits their freedom. Namely, dimensionless vavelength has to be equal to 1 implying $B = 2\,K(m)$ and then $m$ is determined by $\alpha,\beta$ through condition
\begin{equation}\label{lpkdv}
\frac{\alpha}{\beta} = \frac{16}{3}\,m\,K^2(m).
\end{equation}
Fig. \ref{aDb} displays function $\frac{16}{3}\,m\,K^2(m)$.

\begin{figure}[tbh] 
\begin{center}
\resizebox{0.9\columnwidth}{!}{\includegraphics{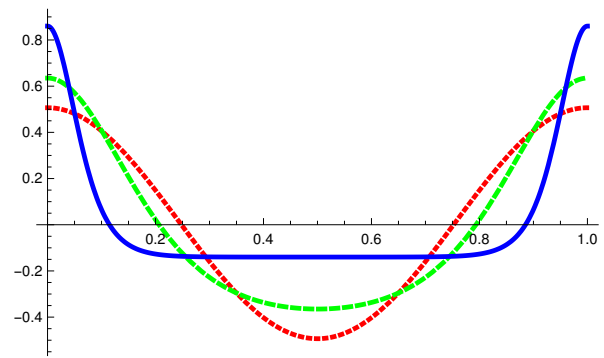}} 
\end{center}
\vspace{-5mm}
\caption{Profiles of cnoidal solutions of KdV for $m=0.1$ - red dotted line, $m=0.9$ - green dashed line and $m=0.99999$ - blue solid line. These values of $m$ correspond to $\alpha/\beta\approx 1.39, 31.9, 272.1$, respectively.} \label{cnKdV}
\end{figure}

Fig. \ref{cnKdV} presents examples of three profiles of cnoidal waves for three different values of the ratio $\alpha/\beta$.

It is clear that if $\beta$ is of the same order than $\alpha$ the cnoidal solution is similar to usual cosine wave. Only when $\beta$ is much smaller than $\alpha$ (2 orders of magnitude smaller) than through shapes become different than those of crests.

The limit $m\to 1$ gives the single soliton solution discussed in previous subsection.

\section{Exact single soliton solution for KdV2}\label{1kdv2}
In \cite{KRI14} we found exact single solution for KdV2 assuming the same form of the solution as for KdV, that is (\ref{asumkdv1}). Below we briefly remind that result. Insertion of (\ref{asumkdv1}) into (\ref{kdv2}) gives (after some simplifications) equation analogous to (\ref{akdv})
\begin{equation}\label{akdv2}
C_0 + C_2 \sech^2(By)+ C_4 \sech^4(By)=0
\end{equation}
which supplies three conditions on parameters of solution formula
\begin{align} \label{C0}
C_0 = & (1-v) + \frac{2}{3}  B^2 \beta + \frac{38}{45}  B^4 \beta^2, \\ \label{C2}
C_2 = &  \frac{3\, \alpha}{4} -  B^2 \beta + \frac{11}{4} \alpha\, B^2  \beta -  \frac{19}{3} B^4 \beta^2,\\ \label{C4}
C_4 = &  -\left(\frac{1}{8}\right)  \alpha^2 - \frac{43}{12} \alpha\, B^2  \beta +  \frac{19}{3}  B^4 \beta^2.
\end{align} 
From (\ref{C4}), denoting $\displaystyle z=\frac{\beta B^2}{\alpha }$ we obtain
\begin{equation} \label{rkw}
\frac{19}{3} z^2 - \frac{43}{12} z -\frac{1}{8}=0
\end{equation}
with roots
\begin{equation} \label{rkw1}
\begin{array}{llll} z_1 & = &\displaystyle \frac{43-\sqrt{2305}}{152}\approx -0.033 & <0\\
 z_2 &= & \displaystyle \frac{43+\sqrt{2305}}{152}\approx 0.599 & > 0. \end{array}
\end{equation}
Thus $B$ is real only when $z=z_2$,
\begin{equation} \label{wB}
B^2 = \frac{\alpha}{\beta}\, z_{2}=  \frac{\alpha}{\beta}\,\frac{43+\sqrt{2305}}{152} \approx 0.599\, \frac{\alpha}{\beta}.
\end{equation}
Equation (\ref{C2}) in consistent with (\ref{wB}) only when
\begin{equation} \label{alfB}
\alpha=\alpha_s = \frac{3(51-\sqrt{2305})}{37} \approx 0.242399.
\end{equation}
Then  (\ref{C0}) determines velocity
\begin{equation} \label{velB}
v=1 +\frac{2}{3} \alpha_s z_2+\frac{38}{35}(\alpha_s z_2)^2 \approx 1.114546.
\end{equation}
These results are the same as in \cite[Sec.~4]{KRI14} (if $A=1$ is set there).

\begin{figure}[tbh] 
\begin{center}
\resizebox{0.9\columnwidth}{!}{\includegraphics{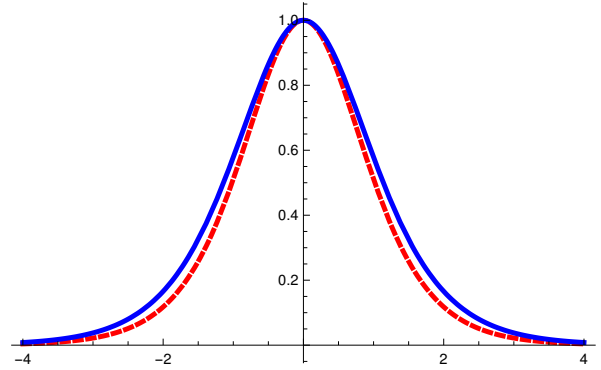}} 
\end{center}
\vspace{-5mm}
\caption{Comparison of the profile of KdV soliton - red dashed line with KdV2 soliton - blue line. Both curves are obtained for the same value of $\alpha=\alpha_s$.} \label{1sol2}
\end{figure}

Comparing single soliton solutions for KdV and KdV2 we see that: 
\begin{itemize}
\item KdV2 sets limitation on parameters of solution. Only one exact solution of KdV2 occurs when $\alpha\approx 0.242399$, whereas solutions of KdV are possible for arbitrary $\alpha$.
\item For KdV2 $\displaystyle B\approx \sqrt{0.6\,\frac{\alpha}{\beta}}$, \hspace{2ex}  for KdV  $\displaystyle B=\sqrt{\frac{3}{4}\frac{\alpha}{\beta}}$. This difference in $B$ values means that the KdV2 soliton is a little wider than that of KdV (for the same $\alpha$), see Fig. \ref{1sol2}.
\item For KdV2 $v \approx 1.114546$, \hspace{2ex} for KdV   $\displaystyle v=1+\frac{\alpha}{2}$.
\item Velocity does not depend on $\beta$ (that is, does not depend on wavelength). For $\alpha \approx 0.242399$ KdV velocity is slightly greater than that of KdV2, with ratio  $\approx 1.006$.
\end{itemize}

\section{Exact periodic solutions for KdV2}\label{pkdv2}

\vspace{-3mm}
We look for periodic nonlinear wave solutions of KdV2 (\ref{kdv2}). 
Introduce $y=x-v t$. Then 
 $\eta(x,t) =\eta(y)$, $\eta_t=-v\eta_y$ and equation (1) takes the form of an ODE
\begin{align} \label{kdv2y}
(1-v) \eta_y & + \frac{3}{2} \alpha\,\eta\eta_y+ \frac{1}{6}\beta\, \eta_{3y} -\frac{3}{8}\alpha^2\eta^2\eta_y \\ & +
  \alpha\beta\,\left(\frac{23}{24}\eta_y\eta_{2y}+\frac{5}{12}\eta\eta_{3y} \right)+\frac{19}{360}\beta^2\eta_{5y} =0 .  \nonumber
\end{align}

Now assume the periodic solution to be in the same form as corresponding  solution of KdV 
\begin{equation} \label{ey}
 \eta(y) = A\, \cn^2(By,m)+D,
\end{equation}
where $A,B,D$ are yet unknown constants ($m$ is the elliptic parameter). The constant $D$ must ensure that the volume of water is the same for all $m$. Here we use the general notation $A$ for the amplitude, later 
we will set $A=1$. [Comparing (\ref{ey}) with (\ref{cn2Ding}) one sees that $ A$ corresponds to $\eta_1 + \eta_2$, $D$ to $-\eta_2$ and $B$ to $1/\Delta$.]


Now we calculate all derivatives $\eta_{ny}$ entering (\ref{kdv2y}). 
Using properties of Jacobi elliptic functions and their derivatives
one can express them as functions of $\cn^2$. So
\begin{align} \label{y1}
\eta_y &= 2AB\,\cn \,[-\sn\dn] = -2B\,\cn\sn\dn, \\ \label{y2a}
\eta_{2y} &= 2AB^2 [1-m + (4m-2)\cn^2-3m\cn^4] ,\\ \label{y3a}
\eta_{3y} &= 8AB^3 \cn\dn\sn[ 1-2m+3m\cn^2] ,\\ \label{y5a}
\eta_{5y} &= -16AB^5 \cn\dn\sn [(2-17m+17m^2)  \\
 & \hspace{4ex} +(30m-60m^2)\cn^2+45m^2\cn^4  ] . \nonumber
\end{align}
Denote (\ref{kdv2y}) as
\begin{equation} \label{eq2}
E_1+E_2+E_3+E_4+E_5+E_6+E_7=0,
\end{equation}
where (common factor $\,\text{CSD}=(-2AB\,\cn\sn\dn)$)
\begin{align} \label{E1}
E_1 &= (1-v) \eta_y =(1-v)\,\text{CSD}, \\ \label{E2}
E_2 &= \frac{3}{2} \alpha\,\eta\eta_y =\frac{3}{2} \alpha\,( \cn^2+D) \,\text{CSD}, 
\\ \label{E3}
E_3 &= \frac{1}{6}\beta\eta_{3x} =-\frac{2}{3}\beta\, B^2 [ 1-2m+3m\cn^2]\,\text{CSD}, 
\\  \label{E4}
E_4 &= -\frac{3}{8}\alpha^2\eta^2\eta_y = -\frac{3}{8}\alpha^2 (\cn^2+D)^2\,\text{CSD}, 
\\ \label{E5}
E_5 &=  \frac{23}{24}\alpha\beta\,\eta_y\eta_{2y}  \\ &=
\frac{23}{12}\alpha\beta\,B^2 [1-m + (4m-2)\cn^2 
 - 3m\cn^4] \,\text{CSD}, \nonumber
 \\ \label{E6}
 E_6 &= \frac{5}{12}\alpha\beta\,\eta\eta_{3y}\\ & =  -  \frac{5}{3}\alpha\beta\,B^2(\cn^2+D)[ 1-2m+3m\cn^2] \,\text{CSD}, \nonumber
 \\ \label{E7}
 E_7&= \frac{19}{360}\beta^2\eta_{5x} =\frac{19}{45}\beta^2 B^4 [(2-17m+17m^2) 
\\ & \hspace{2ex}
+(30m-60m ^2)\cn^2  +45m^2\cn^4  ]  \,\text{CSD}.\nonumber 
 \end{align}
Then (\ref{eq2}) becomes 
\begin{equation} \label{eq2a}
(-2B\,\cn\sn\dn)[F_0 +F_1 \cn^2 +F_2 \cn^4] =0.
\end{equation}
Equation (\ref{eq2a}) is valid for arbitrary argument of $\cn^2$ when all three coefficients $F_0,F_1,F_2$ vanish simultaneously. This gives us a set of three equations for the coefficients $v,B,D$ 
\begin{align} \label{c1}
F_0 &=690 \alpha A   \beta  B^2 (m\!-\!1) \!-\!(\beta B^2)^2 (2584\,m(m\!-\!1)+304)
\nonumber \\ & \hspace{2ex} +240 \beta B^2(1-2m) -60 \alpha  D \left(10 \beta  B^2 (2  m-1)+9\right) \nonumber \\& \hspace{2ex} +135 (\alpha D)^2+360 (v-1) =0, \\ \label{c2}
F_1 &= 90 \alpha  A \left[22 \beta  B^2 (1-2 m)+3 \alpha  D-6\right]  \\ &  \hspace{2ex} +120 \beta  B^2 m
   \left[ 38 \beta  B^2 (2 m-1)+15 \alpha  D+6\right] =0, \nonumber
\\ \label{c3}
F_2 & = 45 \left(3 \alpha ^2 A^2 \!+\!86 \alpha A  \beta  B^2 m\!-\!152 \beta ^2 B^4 m^2\right)=0.
\end{align}
Equations (\ref{c1})-(\ref{c3}), supplemented by the volume conservation law, allow us to find all unknowns as functions of the elliptic parameter $m$.
Below we show these solutions explicitly.
\begin{equation} \label{z_def}  \mbox{Now,~denote} \hspace{13ex}
z = \frac{B^2\,\beta}{A\,\alpha} m. \hspace{10ex}
\end{equation}
Then, equation (\ref{c3}) becomes identical with (\ref{rkw}) and has the same roots (\ref{rkw1}).

\subsection{Volume conservation}

In principle, exact periodic solutions of KdV2 with $D=0$ exist. They make sense from a mathematical point of view. For KdV case the derivation of such periodic solutions is presented in Whitham's book \cite{Whit}. The more careful derivation, presented by Dingemans \cite{Ding}, stresses that periodic solutions should have profile uplifts and depressions with respect to the undisturbed water level. Therefore the volume conservation condition is crucial for obtaining proper physical solutions.

Volume conservation determines the value of $D$. Here by mass conservation we mean that  each  $m$ solution involves the same volume of water
\begin{equation} \label{vcons}
\int_0^L (A\cn^2(By,m)+D)\,dy =0. \nonumber 
\end{equation} Then
\begin{equation} \label{vcons1}
D=-\frac{A}{L} \int_0^L \cn^2(By,m)\,dy  \equiv -\frac{A}{L}\, I(L)\,,
\end{equation}
where $L=$~ is the wavelength.
The periodicity condition implies
\begin{equation} \label{lam}
\cn^2\left(Bl,m\right)=\cn^2(0,m) \quad \Longrightarrow \quad  L=\frac{2\,K(m)}{B} ,
\end{equation}
where $K(m)$ is the  complete elliptic integral of the first kind. Hence
\begin{equation} \label{dj}
D= \!-\frac{A}{L}I(L)\! = \!-\frac{ [E(\text{am}(2 K(m)|m)|m)\!+\!\ (m\!-\!1) K(m)]}{2 m K(m)}, 
\end{equation}
where ~$E(\Theta|m)$ is the elliptic integral of the second kind and ~$\textrm{am}(x|m)$ is  the Jacobi elliptic function amplitude.
Since 
\begin{equation} \label{EEm}
\frac{E(\textrm{am}(2K(m)|m)|m))}{2\,K(m)}\equiv \frac{E(m)}{K(m)},
\end{equation} 
where $E(m)$ is the complete elliptic integral, and (\ref{dj}) simplifies to 
\begin{equation} \label{ddd}
D=-\frac{A}{m}\left[ \frac{E(m)}{K(m)} + m-1\right].
\end{equation}
The function 
$\left[\frac{E(m)}{K(m)} + m-1\right]$ is positive for $m\in (0,1)$ and vanishes at $m=0$ and $m=1$.

\subsection{Coefficients of the exact solutions to KdV2} \label{ss2}

Without any assumptions on $m,\alpha,\beta$,~ other than ~$0\le m\le 1$~ 
we obtained the set of four conditions (\ref{c1})-(\ref{c3}) and (\ref{ddd}) on $A,B,D,v$ and $m$. Since
equation  (\ref{c3}) admits two values for $z$ then we have to consider 
 two different cases. 

\subsubsection{Case ~$\displaystyle z=z_2=\frac{43+\sqrt{2305}}{152}$~ }

Solving the set (\ref{c1})-(\ref{c3}) and (\ref{ddd})
for $z=z_2$ one obtains
\begin{align}\label{A2}A=&\frac{3 \left(-51+\sqrt{2305}\right) m}{37\, \alpha \, \text{EK}(m)} = -\frac{\alpha_s \,m}{\alpha\,\text{EK}(m)},
\\ \label{B2}
B=& \hspace{1ex}\sqrt{-\frac{3 \left(-14+\sqrt{2305}\right)}{703\,\beta\,\text{EK}(m)} },
\\ \label{D2}
D=& \hspace{1ex} 
\frac{ \left(51-\sqrt{2305}\right)
}{37\, \alpha} \left(1+\frac{2m-1}{\text{EK}(m)}\right),
\\ \label{v2}
v = & \hspace{1ex}  \frac{9439-69
   \sqrt{2305}}{5476} \\ & 
-\frac{\left(377197-7811
   \sqrt{2305}\right)
   \left(m^2-m+1\right)}{520220\, \text{EK}(m)^2} \nonumber \\
 \approx & \hspace{1ex}1.11875\, -0.00420523 \,\frac{ (m^2-m+1)}{\text{EK}(m)^2}. \nonumber 
\end{align}
Hence, $B$ is real-valued only when ~$\text{EK}(m)  < 0$,~ 
that is for  
\begin{equation} \label{Mcond}
 m>m_s\approx 0.9611494753812.
\end{equation}
Therefore, for the branch of solutions connected to $z=z_2$, the elliptic parameter $m\in (m_s,1]$.
For $m>m_s$,~  the amplitude ~$A>0$.

Notice that the velocity depends only on $m$. 

\begin{table}[bth] 
\caption{Values of $m,\beta,B,D,v$ as functions of $\alpha$. Branch ~$z=z_2=\frac{43+\sqrt{339}}{152}$.} \label{tab1}
\begin{center}
\begin{tabular}{||c|c|c|c|c|l||} \hline 
~~~~$\alpha$~~~~&~~~~$m$~~~~&~~~~$\beta$~~~~&~~~~$B$~~~~&~~~~$D$~~~~&~~~~$v$~~~~  \\ \hline 
 $\approx$0.242399 &  1  &  0  &  $\infty$  & 0 &   1.11455  \\ \hline
 0.28580  &  1-10$^{-16}$  &  0.000110  &  39.5094  & -0.05062 &   1.11290  \\ \hline
 0.29253  &  1-10$^{-14}$  &  0.000142  &  35.0096  & -0.05713 &   1.11263  \\ \hline
 0.30200  &  1-10$^{-12}$  &  0.000196  &  30.4036  & -0.06578 &   1.11222  \\ \hline
 0.31586  &  1-10$^{-10}$  &  0.000284  &  25.7984  & -0.07752 &   1.11161  \\ \hline
 0.35522  &  1-10$^{-7}$   &  0.000596  &  18.8907  & -0.10587 &   1.10972  \\ \hline
 0.40     &  0.9999961     &  0.001033  &  15.2271  & -0.13134 &   1.10730  \\ \hline
 0.50     &  0.9998598     &  0.002208  &  11.6455  & -0.17169 &   1.10085  \\ \hline
 0.60     &  0.999319      &  0.003548  &  10.0659  & -0.19845 &   1.09296  \\ \hline
\end{tabular}
\end{center}
\end{table}

Now, we use conditions that in dimensionless variables $A=1$ and $L=1$.
Then, (\ref{A2}) gives 
\begin{equation} \label{alfz2}
\alpha = - \alpha_s \frac{m}{\text{EK}(m)}
\end{equation}
and (\ref{lam}), that is, $B= 2K(m)$ yields
\begin{align} \label{betz2}
\beta & =-\frac{\alpha_s\,z_2}{4} \,\frac{1}{ K^2(m)\,\text{EK}(m)} 
\approx - 0.036\,\frac{1}{ K^2(m)\,\text{EK}(m)}. 
\end{align}
Thus the ratio $\alpha/\beta$ for solutions of KdV2 is
\begin{equation} \label{albez2}
\frac{\alpha}{\beta} \approx 6.7\, m\,K^2(m).
\end{equation}
The relation (\ref{albez2}) differs from that for KdV (\ref{lpkdv}) only by small factor (6.7 instead of 5.33).

We can discuss results with respect to $\alpha$. Given $\alpha$ determines $m$ by (\ref{alfz2}), then (\ref{betz2}) fixes $\beta$ and next (\ref{lam}), (\ref{D2}) and (\ref{v2}) supply values of $B,D$ and~$v$. From (\ref{alfz2}) the smallest possible $\alpha$ is obtained for $m=1$. Its numerical value is $\alpha(m=1)\approx 0.242399$. 
Table \ref{tab1} gives several values of parametrs $m,\beta,B,D,v$  for $\alpha\le 0.6$. In all presented cases velocity is positive and greater than one. This means that waves move toward right both in fixed frame and in the frame moving with natural velocity 1 ($\sqrt{gh}$ in dimension variables).

\begin{figure}[tbh]\begin{center}\resizebox{0.99\columnwidth}{!}{\includegraphics{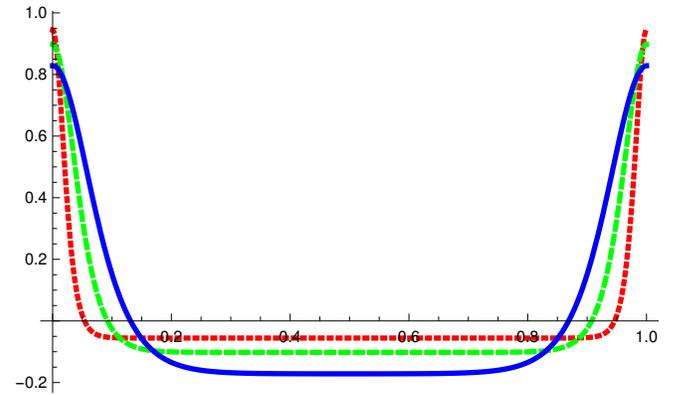}}\end{center}\vspace{-5mm}\caption{Case $z=z_2$.  Profiles for cnoidal waves for $\alpha=0.26$ - red dotted line, $\alpha=0.35$ - green dashed line and $\alpha=0.5$ - blue line.} \label{cnoidy_z2}\end{figure}

Examples of four profiles of cnoidal waves according to parameters contained in table \ref{tab1} are displayed in Fig.~\ref{cnoidy_z2}.

\begin{figure}[tbh] 
\begin{center}
\resizebox{0.9\columnwidth}{!}{\includegraphics{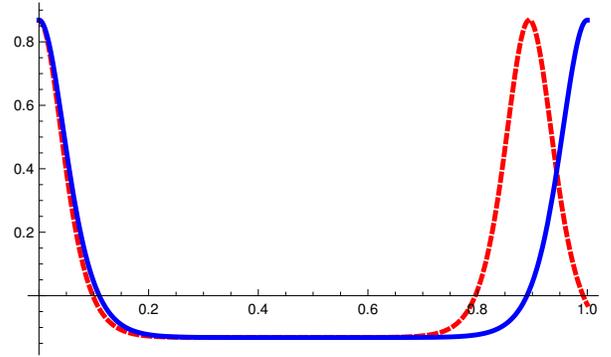}} 
\end{center}
\vspace{-5mm}
\caption{Comparison of the profile of KdV cnoidal wave - red dashed line with KdV2 one - blue line. Both curves are obtained for the same value of $\alpha=0.4$.} \label{cn2sol2}
\end{figure}
In Fig. \ref{cn2sol2} we show a comparison of profiles of periodic  solutions of KdV and KdV2. Both curves correspond to solutions obtained with the same $\alpha=0.4$. The parameters of solutions are: $B\!=\!15.2018, D\!=\!-0.1317$ for KdV2 solution and  $B\!=\!17.2005, D\!=\!-0.1315$ for KdV solution. Since $B_{\text{kdv}}>B_{\text{kdv2}}$, KdV wave is shorter.

\subsubsection{Case ~$\displaystyle z=z_1=\frac{43-\sqrt{2305}}{152}$~ }

Now, 
\begin{align}\label{A1}
A=&-\frac{3 \left(51+\sqrt{2305}\right) m}{37\, \alpha \, \text{EK}(m)}
\approx -8.02787 \,\frac{m}{\alpha\,\text{EK}(m)}
\\ \label{B1}
B=& \hspace{1ex}\sqrt{\frac{3 \left(14+\sqrt{2305}\right)}{703\,\beta\,\text{EK}(m)} }
\\ \label{D1}
D=&  \hspace{1ex}
\frac{ \left(51+\sqrt{2305}\right)
}{37\, \alpha} \left(1+\frac{2m-1}{\text{EK}(m)}\right), \\ \label{v1} 
v = &\hspace{1ex} \frac{9439+69  \sqrt{2305}}{5476} \\& 
-\frac{\left(377197+7811  \sqrt{2305}\right)
   \left(m^2-m+1\right)}{520220\, \text{EK}(m)^2} \nonumber\\
 \approx & \hspace{1ex} 2.32866\, -1.44594\,\frac{
   \left(m^2-m+1\right)}{\text{EK}(m)^2} \nonumber
\end{align}

In this case we set $|A|=1$ which gives $m$ as function of  $\alpha$. Then
$L=1$, that is $B=2 K(m)$ determines $\beta$. Next values $B,D,v$ are 
given by (\ref{B1})-(\ref{v1}). Collection of these parameters for several values of $\alpha\le 0.6$ is presented in Table \ref{tab2}.
\begin{table}[bth] 
\caption{Values of $m,\beta,B,D,v$ as functions of $\alpha$. Branch ~$z=z_1=\frac{43-\sqrt{339}}{152}$.} \label{tab2}
\begin{center}
\begin{tabular}{||c|c|c|c|c|l||} \hline 
~~~~$\alpha$~~~~&~~~~$m$~~~~&~~~~$\beta$~~~~&~~~~$B$~~~~&~~~~$D$~~~~&~~~~$v$~~~~  \\ \hline 
 0.01     &  0.001245      &  0.026812  &  3.14257  & 0.49992 &   0.88266  \\ \hline
 0.10     &  0.012379      &  0.026813  &  3.15138  & 0.49922 &   0.88241  \\ \hline
 0.20     &  0.024604      &  0.026814  &  3.16119  & 0.49844 &   0.88165  \\ \hline
 0.30     &  0.036675      &  0.026816  &  3.17101  & 0.49766 &   0.88039  \\ \hline
 0.40     &  0.048593      &  0.026818  &  3.18084  & 0.49689 &   0.87862  \\ \hline
 0.50     &  0.060359      &  0.026822  &  3.19068  & 0.49611 &   0.87641  \\ \hline
 0.60     &  0.071975      &  0.026826  &  3.20053  & 0.49533 &   0.87357  \\ \hline
\end{tabular}
\end{center}
\end{table} 

In this case the ratio $\alpha/\beta$ is 
\begin{equation} \label{albez1}
\frac{\alpha}{\beta} =\frac{4(43+\sqrt{2305})}{3}\, m\,K^2(m) \approx 121.35\, m\,K^2(m)
\end{equation}

It is worth to note that:
\begin{itemize}
\item In this branch, since $A=-1$, solutions are "inverted" cnoidal functions (with crests down and troughs up). This is completely new result.
\item Since $\alpha$ should be small KdV2 admits solutions only in narrow interval of $m$ close to 0. Then shapes of these "inverted" cnoidal functions are very close to shapes of usual cosine functions.
\item All such solutions are unidirectional waves, since in fixed reference frame ($0<v<1$). However, when motion is considered in a frame moving with with a natural velocity (here equal 1) then their velocity becomes negative. It is different from the case of usual cnoidal solutions obtained for the other branch (with $z=z_2$) which have $v>1$ in fixed reference frame.
\end{itemize}
\vspace{-2mm}

\section{Numerical evolution} \label{NumEv}

In  order to check our analytic results  we numerically followed the evolution of several cnoidal waves. We used the finite difference (FDM) code developed for KdV2 in fixed frame (\ref{kdv2}) in our previous papers \cite{KRR14,KRI14}. 
 In examples presented in this subsection we assume the initial wave to be the exact cnoidal wave $\eta(x,t) =  \cn^2[B(x-vt),m]+D$, where $\beta,B,D,v,m$ were taken for a given $\alpha$ from  Table~I 
in the case of normal cnoidal wave or from Table~II in the case of inverted ones.
 The algorithm used was the Zabusky-Kruskal one \cite{ZK}, modified in order to include additional terms. The space derivatives of $\eta(x,t)$ were calculated numerically step by step from the grid values of the function and lower order derivatives by a nine-point central difference formula.
Calculations were performed on the interval $x\in[0,1]$ with periodic boundary conditions of $N$ grid points.  The time step $\Delta t$ was chosen as in \cite{ZK}, i.e.,  $\Delta t\le (\Delta x)^3/4$. The calculations shown in this paper used grids with $N=200$. In calculations presented below
the number of time steps reached $10^7-10^8$. In all cases the algorithm secures  volume (mass) conservation up to 10-11 decimal digits. The precision of our model was confirmed in our studies with the finite element method (FDM) \cite{KRSB_amcs,KRSB_cmst}.
\vspace{-2mm}

\begin{figure}[tbh] \begin{center}\resizebox{1.0\columnwidth}{!}{\includegraphics[angle=-90]{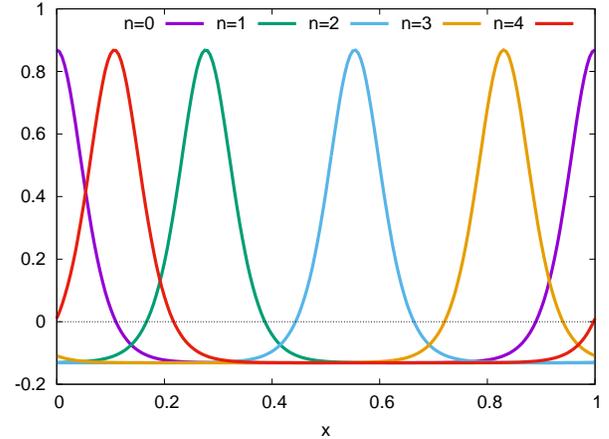}}  \end{center}\vspace{-5mm}\caption{Time evolution of the normal cnoidal wave for the case $\alpha=0.4$. Profiles are displayed at time instants $t_n=n \, dt$, where $dt=\frac{1}{4}$.} \label{cn2z2al04}\end{figure}

Example of motion of normal cnoidal wave, solution of KdV2 equation, obtained with numerical evolution for $\alpha=0.4$ is shown in Fig.~\ref{cn2z2al04}. Numerical solution is stable for much longer time intervals, as well.

The same stability of profiles of numerical solutions is  obtained for inverted cnoidal waves, with parameters listed in Table~II. We do not present them since their profiles are within line thickness almost the same as a cosine function.
\vspace{-5mm}

\section{Conclusions} 
\vspace{-4mm}

From our study the following conclusions can be drawn:
\begin{enumerate}
\item For exended Korteweg - de Vries equation exact solutions, both solitonic and periodic exist. These solutions have the same form as corresonding solutions of KdV equations but with coefficients altered.
\item  KdV2 equation imposes severe limitations on its exact solutions. 
For solitonic case the KdV2 solitons can occur for only one value of $\alpha=\alpha_s$, that is only particular ratio of the wave amplitude to water depth. It might be a reason that solitary waves on the water surface are observed rarely. Physically relevant periodic solutions of KdV2 are related to two narrow intervals of $m$ parameter. For $m$ very close to 1, normal cnoidal waves are obtained. For $m$ very close to 0, inverted cnoidal waves are found. In this case, however, wave profiles are  hardly distinguishable from a cosine function.
\end{enumerate}
\vspace{-5mm}

\begin{acknowledgments}
We would like to thank the anonymous referees for their questions and comments which helped us  formulate the final form of the paper.

One of the authors, GR, would like to thank the National Centre for Nuclear Research, Warsaw, and the University of Zielona G\'ora, for
his stay in 2016.
\end{acknowledgments}
\vspace{-2mm}


\end{document}